\documentclass[twocolumn,prd,10pt,nofootinbib,superscriptaddress]{revtex4-1}
\usepackage{}
\usepackage{epsfig,latexsym,amssymb}
\usepackage{latexsym}
\usepackage{amsmath}
\usepackage{amssymb}
\usepackage{amsfonts}
\usepackage{comment}
\usepackage{graphicx}
\usepackage{verbatim}
\usepackage{epstopdf}
\usepackage{hyperref, url}
\hypersetup{colorlinks   = true,
            urlcolor     = blue,
            citecolor    = blue,
            linkcolor    = blue,
            menucolor    = blue,
            anchorcolor  = blue,
            filecolor    = blue}

\newcommand{\be}{\begin{equation}}
\newcommand{\ee}{\end{equation}}
\newcommand{\bea}{\begin{eqnarray}}
\newcommand{\eea}{\end{eqnarray}}

\begin{document}

\title{Novel standard candle: Collapsing axion stars}

\author{Haoran Di}\email[Corresponding author: ]{hrdi@ecut.edu.cn}
\affiliation{School of Science, East China University of Technology, Nanchang 330013, China}
\author{Lijing Shao}
\affiliation{Kavli Institute for Astronomy and Astrophysics, Peking University,
Beijing 100871, China}
\affiliation{National Astronomical Observatories, Chinese Academy of Sciences,
Beijing 100012, China}
\author{Zhu Yi}
\affiliation{Faculty of Arts and Sciences, Beijing Normal University, Zhuhai 519087, China}
\affiliation{Advanced Institute of Natural Sciences, Beijing Normal University, Zhuhai 519087, China}
\author{Shi-Bei Kong}
\affiliation{School of Science, East China University of Technology, Nanchang 330013, China}
\begin{abstract}
The Hubble constant, $H_0$, is a crucial parameter in cosmology. However, various cosmic observations have produced differing posterior values for $H_0$, resulting in what is referred to as the $H_0$ tension. To resolve this discrepancy, utilizing other cosmological probes to constrain $H_0$ is advantageous. In the quest to identify dark matter candidates, the QCD axion and axionlike particles, collectively referred to as axions, have become leading contenders. These elusive particles can coalesce into dense structures known as axion stars via Bose-Einstein condensation. When these axion stars exceed a critical mass, typically through accretion or merging, they experience a self-induced collapse.
This process results in short radio bursts, assuming a decay constant $f_a\lesssim10^{13}{\rm{GeV}}$, with the frequency depending on the axion mass and the luminosity determined by both the axion mass and decay constant.  Therefore, we propose that collapsing axion stars could serve as a novel standard candle to constrain $H_0$. Even more interesting is that the radio bursts emitted by collapsing axion stars with specific parameters match the characteristics of observed non-repeating fast radio bursts (FRBs). Thus, FRBs generated by collapsing axion stars have the potential to be used as standard candles to constrain $H_0$.
\end{abstract}

\maketitle

\section{Introduction}
The standard cosmological model, known as the $\Lambda$ cold dark matter ($\Lambda$CDM) model, is the simplest framework that aligns well with observational data. Notably, the Planck satellite mission's observations \cite{Planck:2015fie} strongly support a basic 6-parameter $\Lambda$CDM cosmology. However, cracks are beginning to appear in this model, as evidenced by tensions between different astrophysical observations that rely on the 6-parameter $\Lambda$CDM framework. For instance, the standard $\Lambda$CDM model, informed by Planck observations of the cosmic microwave background (CMB) power spectra, predicts a lower Hubble constant ($H_0$) than what is measured locally using the distance ladder method. The $H_0$ value obtained from Cepheid-calibrated Type Ia supernovae (SNe Ia) is $73.04\pm1.04~\rm{km~s^{-1}~Mpc^{-1}}$ \cite{Riess:2021jrx}, which is 5$\sigma$ higher than $67.4\pm0.5~\rm{km~s^{-1}~Mpc^{-1}}$ \cite{Planck:2018vyg} estimated from Planck CMB data within the $\Lambda$CDM framework.
Numerous efforts have been made to resolve the so-called ``Hubble tension'', but none have provided a convincing explanation so far (see Ref. \cite{Feeney:2018mkj} and references therein). This underscores the need to develop new, precise cosmological probes for cross-check.
The gravitational-wave (GW) standard siren method \cite{Schutz:1986gp,Holz:2005df}  is among the most promising approaches to address the Hubble tension. The unique multi-messenger observation, GW170817, provided the first $H_0$ measurement using the standard siren method, achieving about $14\%$ precision \cite{LIGOScientific:2017adf}, which is still insufficient to resolve the Hubble tension.

Another challenge for the $\Lambda$CDM model is the unresolved nature of cold dark matter. The QCD axion \cite{Weinberg:1977ma,Wilczek:1977pj}, which arises from the Peccei-Quinn mechanism \cite{Peccei:1977ur,Peccei:1977hh}, is one of the leading dark matter candidates, providing a prominent solution to the strong-CP problem.
Additionally, string theory strongly suggests the existence of a wide range of axionlike particles (ALPs) across various mass scales, leading to the concept of the ``axiverse" \cite{Arvanitaki:2009fg}. For brevity, we will refer to both the QCD axion and ALPs as ``axion" throughout this paper. Axions, which can emerge through various mechanisms \cite{Preskill:1982cy,Abbott:1982af,Dine:1982ah,Gorghetto:2020qws,Co:2019jts}, have the potential to attain extremely high phase space density, leading to the intriguing phenomenon of Bose-Einstein condensation (BEC) \cite{Sikivie:2009qn} due to their bosonic nature. This condensation enables axions to coalesce into gravitationally bound structures known as axion stars \cite{Braaten:2019knj,Visinelli:2021uve,Gorghetto:2024vnp,Chang:2024fol}.
In the conventional post-inflationary scenario, where the $U(1)_{\rm{PQ}}$ symmetry undergoes spontaneous breaking after the inflationary period, axion stars could potentially make up as much as $75\%$ of the dark matter component \cite{Eggemeier:2019khm,Xiao:2021nkb,Eggemeier:2022hqa}. Moreover, studies of microlensing events from HSC and OGLE data consistently suggest that around $27^{+7}_{-13}\%$ of dark matter may exist in the form of axion stars \cite{Sugiyama:2021xqg}. The observed anomalies in the orbits of trans-Neptunian objects could be explained by the presence of an axion star captured within the solar system \cite{Di:2023xaw}. Axion stars for ALPs with specific parameters may undergo collapse, emitting millisecond-long radio bursts with a peak luminosity of $1.60\times10^{42}\rm{erg/s}$, consistent with the characteristics of observed non-repeating fast radio bursts (FRBs) \cite{Di:2023nnb}. Other studies have also explored the connection between axion stars and FRBs \cite{Iwazaki:2014wka,Tkachev:2014dpa,Raby:2016deh,Buckley:2020fmh}.
The astronomical investigation of axion stars can complement laboratory searches for axions.

This paper is organized as follows. In section II, we introduce another cosmological probe, FRBs, and discuss their limitations. In section III, we briefly present the concept of the dilute axion star. In section IV, we explore the stimulated decay \cite{Di:2023nnb,Kephart:1994uy,Rosa:2017ury,Caputo:2018vmy} of collapsing axion star and explain why this process is suitable for use as a novel standard candle. In section V, we propose that FRBs generated by collapsing axion stars can serve as potential standard candles. The conclusion is presented in section VI. Throughout this paper, we use natural units with $c = \hbar=1$.

\section{Fast radio bursts}
FRBs are intense, transient radio signals that last only a few milliseconds \cite{Lorimer:2007,Keane:2012yh,Thornton:2013iua}, and both their physical origin and the mechanism behind their emission remain largely enigmatic. To date, about a thousand FRBs have been observed, with a small subset exhibiting repeating patterns \cite{Spitler:2016dmz,CHIMEFRB:2019pgo,Fonseca:2020cdd,CHIMEFRB:2023myn}. It is widely believed that repeating and non-repeating FRBs are produced by different physical processes. Numerous models have been proposed to explain the sources of FRBs (see Refs. \cite{Katz:2018xiu,Popov:2018hkz,Platts:2018hiy,Petroff:2019tty,Cordes:2019cmq,Zhang:2020qgp,Xiao:2021omr} for reviews). The frequency spectrum of FRBs typically ranges from about $400\rm{MHz}$ to $8\rm{GHz}$ \cite{Petroff:2019tty}, with total emitted energies between $10^{38}$ and $10^{40}\rm{erg}$ \cite{Thornton:2013iua,Spitler:2014fla}. While there is a confirmed association between at least some FRBs and magnetars  \cite{CHIMEFRB:2020abu,Bochenek:2020zxn}, the exact mechanisms that trigger these mysterious events and their radiation processes remain hotly debated. Similar to the diversity seen in supernovae, it is possible that the observed FRB events originate from a mix of different populations or progenitor mechanisms. Despite these uncertainties, FRBs hold great potential as valuable cosmological probes.

As FRB photons travel through plasma, they interact with free electrons, causing dispersion across different frequencies. The Dispersion Measure is defined as the integral of the free electron number density along the propagation path, and it is directly proportional to the cosmological distance. This makes Dispersion Measure a useful tool for measuring cosmological distances. If an FRB can be precisely localized to a specific galaxy, its redshift can be inferred from the host galaxy \cite{Chatterjee:2017dqg,Bannister:2019iju,Niu:2021bnl}. However, accurately determining redshift by identifying the burst's host galaxy has proven difficult, as the contributions to Dispersion Measure from the host galaxy and the inhomogeneities of the intergalactic medium cannot be precisely determined from observations. Of the approximately one thousand of FRBs detected, only about thirty have been precisely localized \cite{Wei:2023avr,Arcus:2024pvz}. Despite the limited number of accurately localized events, FRBs have already been utilized to estimate $H_0$ \cite{Macquart:2020lln,Hagstotz:2021jzu,Wu:2021jyk,Zhao:2022yiv}. However, more data are needed to improve the accuracy of these measurements and determine the most reliable value. For a cosmological probe to be precise, a large dataset is necessary to minimize random errors. Consequently, FRBs have the potential to be excellent cosmological probes if a significantly greater number can be localized.

\section{Dilute Axion Stars}
The QCD axion is a pseudoscalar boson with spin 0, characterized by its small mass $m_\phi$, extremely weak self-interaction, and very feeble interactions with Standard Model particles.
To ensure Lagrangian shift symmetry invariance, the axion potential $V(\phi)$ must exhibit periodic behavior with respect to $\phi$: $V(\phi)=V(\phi+2\pi f_a)$,
where $f_a$ is the axion decay constant, representing the energy scale at which the $U(1)_{\rm{PQ}}$ symmetry spontaneously breaks. The instanton potential \cite{Peccei:1977ur} is the most commonly used model for the axion potential in phenomenological studies:
\bea
V(\phi)&=&m_\phi^2 f_a^2\left[1-\cos\left(\frac{\phi}{f_a}\right)\right]\nonumber\\
&=&{1\over2} m_\phi^2 \phi^2+\frac{\lambda}{4!}\phi^4+...,
\eea
where $\lambda=-m_\phi^2/f_a^2$ is the attractive self-interaction coupling constant.
As bosonic particles, axions can reach very high phase space densities, enabling the formation of BECs. These BECs can give rise to axion stars, which may exist in both dilute and dense forms \cite{Schiappacasse:2017ham,Chavanis:2017loo,Visinelli:2017ooc,Eby:2019ntd}. However, dense axion stars may have lifespans too short to be of significant cosmological relevance as astrophysical objects \cite{Braaten:2019knj,Seidel:1991zh,Hertzberg:2010yz,Eby:2015hyx,Wang:2020zur}.

A stable dilute axion star can be described as a system where the attractive self-gravity of the axions is counterbalanced by the repulsive gradient energy. This equilibrium is maintained as long as the star's density remains low enough that self-interactions are negligible. The maximum mass and corresponding minimum radius of a dilute axion star can be expressed as follows \cite{Chavanis:2011zi,Chavanis:2011zm}:
\bea
\label{maximum mass}
M_{\rm{max}}\sim 5.073\frac{M_{pl}}{\sqrt {|\lambda|}},~~~~~~R_{\rm{min}}\sim\sqrt {|\lambda|}\frac{M_{pl}}{m_\phi}\lambda_c,
\eea
where $M_{pl}$ represents the Planck mass and $\lambda_c$ denotes the Compton wavelength of the axion.
When the axion star's mass increases and surpasses the maximum mass specified by Eq.\eqref{maximum mass} due to merger events \cite{Mundim:2010hi,Cotner:2016aaq,Schwabe:2016rze,Eby:2017xaw,Hertzberg:2020dbk,Du:2023jxh} or the accretion of axions from the surrounding environment \cite{Chen:2020cef,Chan:2022bkz,Dmitriev:2023ipv}, the attractive self-gravity and self-interaction of the axions will exceed the repulsive gradient energy, leading to the collapse of the axion star. As the star begins to collapse, its density increases rapidly. When the collapsing axion star reaches a critical radius, the stimulated decay of the star can be triggered, generating short radio bursts, provided the decay constant $f_a<1.08\times10^{13}{\rm{GeV}}$ \cite{Di:2023nnb}. For larger decay constants, when the radius of an axion star that has exceeded its maximum mass contracts to the axion's Compton wavelength, $2\pi/m_\phi$, which is greater than the critical radius triggering stimulated decay, the axions begin to annihilate and transition into relativistic states \cite{Levkov:2016rkk}. This leads to a rapid loss of the axion star's energy, a phenomenon known as a ``bosenova'' \cite{Chavanis:2016dab,Levkov:2016rkk,Eby:2016cnq,Fox:2023xgx}.

\section{A novel standard candle: collapsing axion stars}
In the framework of the instanton potential, the general Lagrangian for the axion can be written as follows:
\bea \label{axion}
{\cal L}&=&{1\over 2}\partial_{\mu}\phi\partial^{\mu}\phi-{1\over 2} m_{\phi}^2\phi^2
-\frac{\lambda}{4!}\phi^4 \nonumber\\
&&+{1 \over4}g_{a\gamma\gamma}\phi F_{\mu\nu} \tilde F^{\mu\nu}+...,
\eea
where $F_{\mu\nu}$ represents the electromagnetic field tensor, and $\tilde F^{\mu\nu}$ is its dual tensor, defined as $ F_{\alpha\beta} \epsilon^{\mu\nu\alpha\beta}/2$. The axion-photon coupling constant, $g_{a\gamma\gamma}$, is defined as $g_{a\gamma\gamma}=\alpha K/(2\pi f_a)$, where $\alpha$ represents the fine structure constant, and $K$ is a model-dependent constant typically around one. For example, in the standard  Kim-Shifman-Vainshtein-Zakharov (KSVZ) model \cite{Kim:1979if,Shifman:1979if}, $K$ is approximately $-1.95$, while in the Dine-Fishler-Srednicki-Zhitnitskii (DFSZ) model \cite{Dine:1981rt,Zhitnitsky:1980tq}, $K$ is around $0.72$. For simplicity, we assume $K=1$ in the following discussion. For QCD axions, their interaction with gluons establishes a well-known relationship between the decay constant $f_a$ and the axion mass \cite{Sikivie:2006ni}: $m_{\phi}\simeq6\mu{\rm{eV}}\left({10^{12}\rm{GeV}}/{f_a}\right)$.
By substituting the attractive self-interaction coupling constant $\lambda=-m_\phi^2/f_a^2$ and the Compton wavelength of the axion $\lambda_c=2\pi/m_\phi$ into Eq. \eqref{maximum mass},  we obtain the following expressions for the maximum mass and minimum radius of a dilute axion star:
\bea\label{critical mass}
M_{\rm{max}}\sim5.97\times10^{-12}M_{\odot}\left(\frac{m_\phi}{10^{-5}{\rm{eV}}}\right)^{-1}
\left(\frac{f_a}{10^{12}{\rm{GeV}}}\right),\nonumber\\
\eea
\bea\label{radius}
R_{\rm{min}}\sim2.41\times10^2{\rm{km}}\left(\frac{m_\phi}{10^{-5}{\rm{eV}}}\right)^{-1}
\left(\frac{f_a}{10^{12}{\rm{GeV}}}\right)^{-1},\nonumber\\
\eea
where $M_{\odot}$ denotes the solar mass. These equations show that the maximum mass and minimum radius of a dilute axion star depend solely on the axion mass and the decay constant. For more general axion stars, we denote the mass and radius as $M_{\rm{AS}}$ and $R_{\rm{AS}}$, respectively.

Axions are not completely stable, mainly because of their interaction with the electromagnetic field, as described by the interaction term ${\cal L}_{int}=1/4 g_{a\gamma\gamma}\phi F_{\mu\nu} \tilde F^{\mu\nu}$. This interaction causes axions, in their rest frame, to spontaneously decay into two photons with the same helicity due to angular momentum conservation. The decay rate is given by \cite{Kephart:1994uy}
\bea \label{decay rate}
\Gamma_\phi&=&\frac{1}{8\pi}\left(\frac{1}{2m_\phi}\right)
\frac{1}{2}\sum_{\lambda^\prime=\pm}|{\cal M(\phi\rightarrow\gamma(\lambda^\prime)\gamma(\lambda^\prime))}|^2\nonumber\\
&=&1.02\times 10^{-50} {\rm{s}^{-1}} \left( \frac{m_\phi}{10^{-5}\rm{eV}} \right)^3 \left(\frac{10^{12}\rm{GeV}}{f_a}\right)^2,
\eea
where $\lambda^\prime=\pm$ denotes the helicity of the photons, and $\cal M(\phi\rightarrow\gamma(\lambda^\prime)\gamma(\lambda^\prime))$ represents the transition matrix element determined by the interaction term ${\cal L}_{int}=1/4 g_{a\gamma\gamma}\phi F_{\mu\nu} \tilde F^{\mu\nu}$.
Since the axion is a boson, the boson enhancement effect must be considered within the axion star. The evolution of the photon number density with a given helicity $\lambda^\prime=\pm$ within an axion star, driven by axion decays and inverse decays, is described by the Boltzmann equation \cite{Kephart:1994uy}:
\bea
\frac{dn_{\lambda^\prime}}{dt}&=&\int dX_{\rm{LIPS}}|{\cal M(\phi\rightarrow\gamma(\lambda^\prime)\gamma(\lambda^\prime))}|^2\nonumber\\
&&\times\{f_\phi(\textbf{p})[1+f_{\lambda^\prime}(\textbf{k}_1)][1+f_{\lambda^\prime}(\textbf{k}_2)]\nonumber\\
&&-f_{\lambda^\prime}(\textbf{k}_1)f_{\lambda^\prime}(\textbf{k}_2)[1+f_\phi(\textbf{p})]\},
\eea
where $f_\phi$ and $f_{\lambda^\prime}$ represent the phase space densities of axions and photons, respectively. The photon number density,  $n_{\lambda^\prime}$, is calculated as $n_{\lambda^\prime}=\int d^3 k/(2\pi)^3f_{\lambda^\prime}$. The phase space integration is carried out using the standard Lorentz-invariant measure, which accounts for the momenta of both the axion and the photons. To solve this equation, we assume that the phase space distribution of axions and photons within the dilute axion star is roughly homogeneous and isotropic. Under this assumption, the time derivative of total photon number density can be expressed as follows \cite{Rosa:2017ury}:
\bea\label{photon number density}
\frac{dn_\gamma}{dt}=2\Gamma_\phi n_\phi+\frac{16\pi^2}{m_\phi^3v}\Gamma_\phi n_\phi n_\gamma
-\frac{16\pi^2}{3m_\phi^3}
\left(v+\frac{3}{2}\right)\Gamma_\phi n_\gamma^2,\nonumber\\
\eea
where $n_\phi$ represents the axion number density, and $v$ is the maximum velocity of axions within the axion star, estimated to be around $1/(2R_{\rm{AS}}m_\phi)$ based on the Heisenberg uncertainty principle.
In this equation, the first term corresponds to the spontaneous decay of axions. The second term, which is directly proportional to the product of the axion number density and the photon number density, represents the stimulated decay of axions. The final term, proportional to $n_\gamma^2$, accounts for the process of inverse decay. Additionally, the photons generated through spontaneous and stimulated decay escape from the axion star at a rate given by
$\Gamma_e=R_{\rm{AS}}^{-1}$,
which is the inverse of the axion star's radius.
\begin{figure}
\begin{center}
\includegraphics[width=0.45\textwidth]{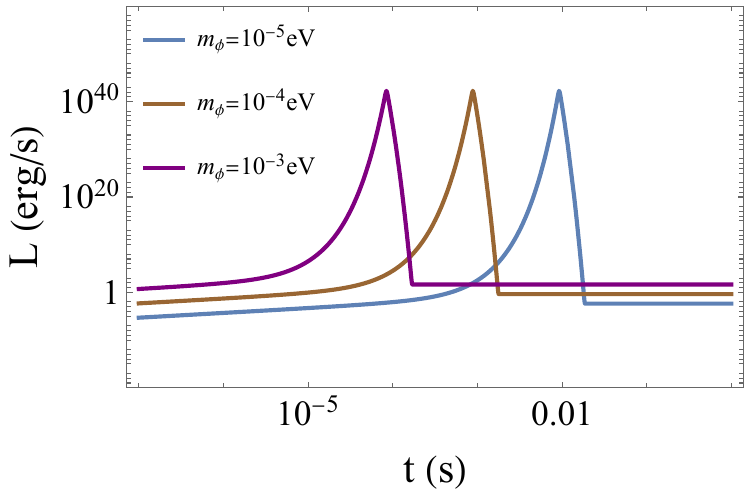}
\caption{Luminosities of collapsing axion stars with a decay constant $f_a=5\times10^{12}{\rm{GeV}}$ and axion masses of $10^{-5}{\rm{eV}}$, $10^{-4}{\rm{eV}}$, and $10^{-3}{\rm{eV}}$, respectively. The peak luminosities of these bursts are approximately $10^{42}\rm{erg/s}$, with the total energy emitted being around $10^{39}\rm{erg}$, $10^{38}\rm{erg}$, and $10^{37}\rm{erg}$, respectively. For a fixed decay constant, the duration of the radio bursts varies depending on the axion mass.}
\label{fig:luminosity}
\end{center}
\end{figure}

\begin{figure}
\begin{center}
\includegraphics[width=0.45\textwidth]{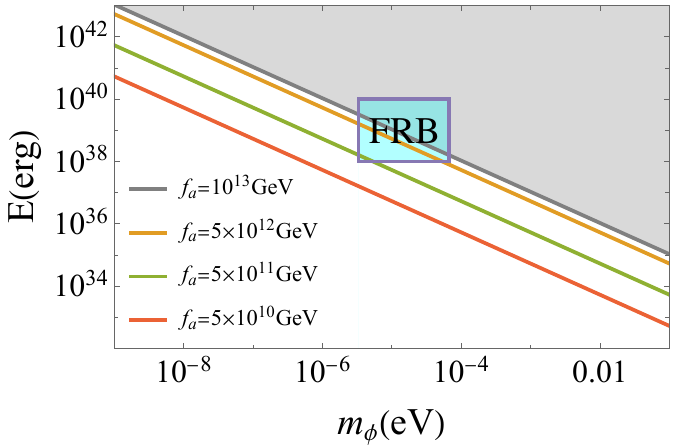}
\caption{Total energy emitted by collapsing axion stars with varying parameters. The total energy depends only on the axion mass and decay constant. With a constant decay constant, the total energy emitted by the collapsing axion star decreases as the axion mass increases. Conversely, with a constant axion mass, the total energy emitted increases as the decay constant rises. The rectangular area corresponds to the luminosity and frequency of currently detected FRBs, with frequency mapped to the axion mass. The gray area indicates where collapsing axion stars would produce relativistic axions rather than radio bursts.}
\label{fig:energy}
\end{center}
\end{figure}

By integrating Eq. \eqref{photon number density} and accounting for photons that escape from the axion star, we derive the following set of coupled differential equations describing the evolution of the photon and axion numbers within the axion star:
\bea \label{coupled differential equation1}
\frac{dN_\gamma}{dt}&=&2\Gamma_\phi N_\phi+\frac{12\pi}{m_\phi^3vR_{\rm{AS}}^3}\Gamma_\phi N_\phi N_\gamma\nonumber\\
&&-\frac{2\pi (2v+3)}{m_\phi^3R_{\rm{AS}}^3}\Gamma_\phi N_\gamma^2-\Gamma_e N_\gamma,
\eea
\bea  \label{coupled differential equation2}
\frac{dN_\phi}{dt}&=&-\Gamma_\phi N_\phi-\frac{6\pi}{m_\phi^3vR_{\rm{AS}}^3}\Gamma_\phi N_\phi N_\gamma\nonumber\\
&&+\frac{4\pi v}{m_\phi^3R_{\rm{AS}}^3} \Gamma_\phi N_\gamma^2.
\eea
For a dilute axion star with a minimum radius, substituting Eq. \eqref{radius} into $\Gamma_e=R_{\rm{AS}}^{-1}$ changes the photon escape rate to
\bea
\Gamma_e\sim1.24\times10^3{\rm{s^{-1}}}\left(\frac{m_\phi}{10^{-5}{\rm{eV}}}\right)
\left(\frac{f_a}{10^{12}{\rm{GeV}}}\right).
\eea
The photon number $N_\gamma\simeq(2\Gamma_\phi/\Gamma_e)N_\phi$ in dilute axion stars with minimum radius, resulting from spontaneous decay, is not sufficient to trigger stimulated decay. However, when the axion star's mass exceeds the critical mass, causing it to collapse until its size becomes smaller than the critical radius
\bea \label{critical radius}
R_{cr}&\simeq&\frac{24\pi\Gamma_\phi M_{\rm{max}}}{m_\phi^3}\nonumber\\
&\sim&6.67\times10^{-4}{\rm{km}}\left(\frac{10^{-5}{\rm{eV}}}{m_\phi}\right)
\left(\frac{10^{12}{\rm{GeV}}}{f_a}\right),
\eea
i.e., when $12\pi\Gamma_\phi N_\phi N_\gamma/(m_\phi^3vR_{\rm{AS}}^3)\simeq2\Gamma_\phi N_\phi$, stimulated decay is initiated.
Once the axion star's radius reaches the critical value, the photon escape rate becomes
\bea \label{escape rate2}
\Gamma_e&\simeq&\frac{m_\phi^3}{24\pi\Gamma_\phi M_{\rm{max}}}\nonumber\\
&\sim&4.50\times10^8{\rm{s^{-1}}}\left(\frac{m_\phi}{10^{-5}{\rm{eV}}}\right)\left(\frac{f_a}{10^{12}{\rm{GeV}}}\right).
\eea
When the star's size approaches the Compton wavelength of the axion, $2\pi/m_\phi$, axions begin to annihilate, transitioning into relativistic states  \cite{Levkov:2016rkk}.
To induce stimulated radiation rather than generating relativistic axions, the critical radius $R_{cr}\simeq24\pi\Gamma_\phi M_{\rm{max}}/m_\phi^3$ must be larger than the Compton wavelength, $2\pi/m_\phi$. This condition requires that $f_a^{-1}\gtrsim10^{-13}\rm{GeV}^{-1}$, as illustrated in Fig.~\ref{fig:energy} and Fig.~\ref{fig:constraints}.

The luminosity of the collapsing axion star due to stimulated decay is given by
\bea \label{luminosity}
L_\phi&=&\frac{1}{2}m_\phi N_\gamma R_{cr}^{-1}=\frac{N_\gamma m_\phi^4}{48\pi\Gamma_\phi M_{\rm{max}}}\nonumber\\
&\sim&3.60\times10^{40}{\rm{erg/s}}\left(\frac{N_\gamma}{10^{49}}\right)
\left(\frac{m_\phi}{10^{-5}{\rm{eV}}}\right)^2\left(\frac{f_a}{10^{12}{\rm{GeV}}}\right),\nonumber\\
\eea
which is determined solely by the mass of the axion and the decay constant $f_a$.
By substituting Eqs. \eqref{decay rate}, \eqref{critical radius} and \eqref{escape rate2} into Eqs. \eqref{coupled differential equation1} and \eqref{coupled differential equation2}, and then numerically solving these coupled equations, we can obtain $N_\gamma$. Substituting this into Eq. \eqref{luminosity} allows us to determine the temporal variation of the collapsing axion star's luminosity, as illustrated in Fig.~\ref{fig:luminosity}. In Fig.~\ref{fig:luminosity}, we present three numerical solutions for a representative set of parameters with a decay constant $f_a=5\times10^{12}{\rm{GeV}}$ and varying axion mass. Once the size of the axion star falls below the critical radius, it will produce a brief radio burst, as shown in Fig.~\ref{fig:luminosity}.
Therefore, a collapsing axion star with an inverse decay constant $f_a^{-1}\gtrsim10^{-13}\rm{GeV}^{-1}$ will trigger stimulated decay.
The spectrum of the radio burst will be nearly monochromatic at a frequency of $\nu\simeq m_\phi/({4\pi})\approx1.21(m_\phi/10^{-5}{\rm{eV}}){\rm{GHz}}$, which is a distinctive signal that could potentially be detected by radio telescopes. If a radio signal consistent with the stimulated decay of a collapsing axion star is observed, it can serve as a standard candle due to the strong luminosity of this radio burst, which is determined solely by the axion mass and the decay constant. Thus, collapsing axion stars could serve as excellent cosmological probes.

The above results are based on the instanton potential of axions. If other potentials are considered, the luminosity and total radiated energy of collapsing axion stars will change. For example, the chiral potential \cite{DiVecchia:1980yfw} has a corresponding coupling constant for the self-interaction of the axion of $-0.34m_\phi^2/f_a^2$ \cite{Fujikura:2021omw}, which is about one-third of the coupling constant for the instanton potential. This will increase the maximum mass of the dilute axion star and raise the critical radius corresponding to the onset of stimulated decay. The maximum luminosity remains unchanged, while the total energy radiated by the axion star increases.
Additionally, the strength of the interaction between axions and photons depends on the selected $K$ values, meaning that different choices for $K$ will affect the luminosity of axion stars. Specifically, the decay rate of an axion is proportional to the square of $K$, rendering the result independent of the sign of $K$. For the KSVZ axion model, $K$ is approximately $-1.95$, with its absolute value being about twice that of the value we consider. Consequently, both the decay rate of the axion and the critical radius will increase, resulting in a decrease in the maximum luminosity of the axion star, while the total energy radiated remains unchanged.

\section{A potential standard candle: ~fast radio bursts}
\begin{figure}
\begin{center}
\includegraphics[width=0.45\textwidth]{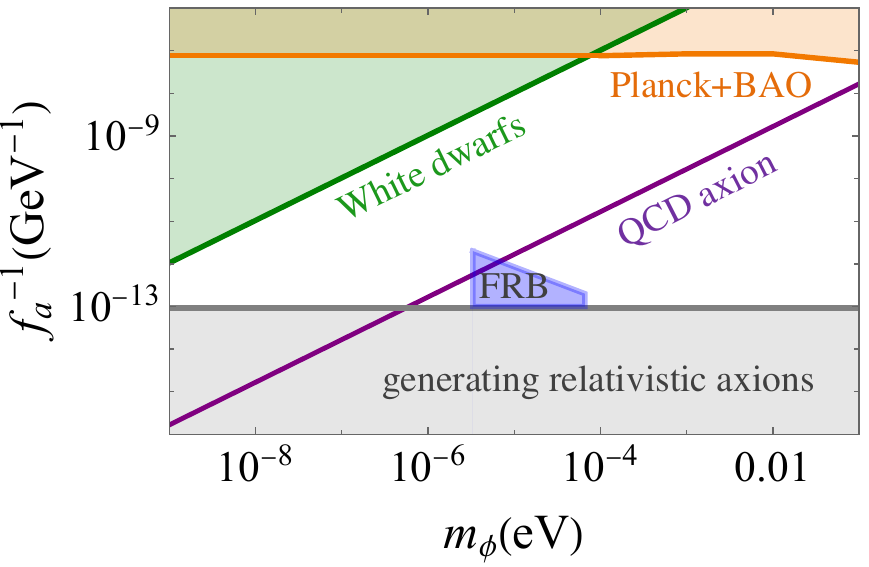}
\caption{Constraints on the axion parameter space. The region above the orange solid line is excluded based on constraints from Planck 2018 data combined with BAO measurements \cite{Caloni:2022uya}. The green area indicates exclusions derived from white dwarf observations \cite{Balkin:2022qer}. The latest constraints on the axion parameter space from white dwarf observations can be found in Ref. \cite{Gomez-Banon:2024oux}. The purple line denotes the parameter space for QCD axions. Below the gray line, collapsing axion stars are predicted to emit relativistic axions \cite{Levkov:2016rkk} rather than radio bursts. The blue region corresponds to parameter space that aligns with the luminosity and frequency characteristics of non-repeating FRBs.}
\label{fig:constraints}
\end{center}
\end{figure}

Interestingly, the radio signals from the stimulated decay of collapsing axion stars may have already been detected. FRBs are bright, transient radio signals lasting just milliseconds, with a frequency spectrum ranging from approximately $400\rm{MHz}$ to $8\rm{GHz}$, and a total energy emission typically between $10^{38}$ and $10^{40}\rm{erg}$. Fig.~\ref{fig:energy} shows the variation of the total energy radiated by a collapsing axion star with different decay constants and axion masses. It can be observed from Fig.~\ref{fig:energy} that the total energy and photon frequency radiated by a collapsing axion star for certain parameters are consistent with some FRBs.
For constraints on the axion parameter space from FRBs, refer to Fig.~\ref{fig:constraints}. Therefore, the stimulated decay of collapsing dilute axion stars might account for some of the observed non-repeating FRBs. Additionally, in the conventional post-inflationary scenario, axion stars could comprise up to $75\%$ of the dark matter component. Given the significant abundance of axion stars in the Universe, collapsing axion stars may be common enough to serve as sources of FRBs. The rate of collapsing axion stars deserves further dedicated study.
The frequency of the observed radio signals is given by
\bea
\nu_{\rm{obs}}=\frac{\nu_{\rm{em}}}{1+z}=\frac{m_\phi}{4\pi(1+z)},
\eea
where $\nu_{\rm{em}}$ is the emitted frequency of the radio signal, determined by the mass of the axion. Therefore, the redshift can be expressed as
\bea
z=\frac{m_\phi}{4\pi\nu_{\rm{obs}}}-1.
\eea
Fig.~\ref{fig:constraints} illustrates the constraints on the axion parameter space based on the hypothesis that some FRBs are explained by collapsing axion stars. This provides a potential direction for experimental axion searches. Once the axion mass is accurately determined through experiments, the redshift of radio bursts can also be precisely measured. The luminosity flux of the collapsing axion star as observed from Earth is given by the formula:
\bea
F_\phi=\frac{L_\phi}{4\pi d_L^2}=\frac{N_\gamma m_\phi^4}{192\pi^2\Gamma_\phi M_{\rm{max}} d_L^2},
\eea
where $d_L$ represents the luminosity distance. The luminosity distance can then be expressed as
\bea
d_L=\left(\frac{L_\phi}{4\pi F_\phi}\right)^{1/2}=\frac{m_\phi^2}{8\pi}
\left(\frac{N_\gamma}{3\Gamma_\phi M_{\rm{max}}F_\phi}\right)^{1/2},
\eea
where $N_\gamma$, $\Gamma_\phi$ and $M_{\rm{max}}$ depend solely on the axion mass and decay constant. Thus, by measuring the flux and knowing the axion mass and decay constant, we can determine the luminosity distance $d_L$. The luminosity distance, measured in this way, provides a cross-check with the cosmological distance obtained from the Dispersion Measure, which is defined as the integral of the number density of free electrons along the propagation path. This cross-check improves the accuracy of the luminosity distance, which in turn can help pinpoint the host galaxy of the FRB. As more FRBs are precisely located, the value of the Hubble constant can be measured with greater accuracy, which may help resolve the Hubble tension. Therefore, FRBs generated by collapsing axion stars have the potential to be used as standard candles and serve as valuable cosmological probes.

\section{Conclusions}
The Hubble constant, $H_0$, is a crucial parameter in cosmology, but different observational methods have produced varying estimates, leading to what is known as the $H_0$ tension. To resolve this discrepancy, it is helpful to employ additional cosmological probes. The GW standard siren method is among the most promising approaches for addressing the Hubble tension. The unique multi-messenger observation event, GW170817, provided the first measurement of $H_0$ using the standard siren method, achieving about 14\% precision, which is not sufficient to resolve the tension. Moreover, FRBs have the potential to be powerful cosmological probes, but the limited number of accurately localized events leads to imprecise measurements of the Hubble constant. If a larger number of FRBs could be localized, they could serve as valuable cosmological probes.

The QCD axion or ALP is a leading dark matter candidate. Axions can collectively form a bound state known as an axion star through BEC. When these axion stars exceed a critical mass due to accretion or merging, they undergo a collapse driven by self-interactions. This collapse results in short radio bursts, with their frequency determined by the axion mass and their luminosity by both the axion mass and decay constant. The resulting radio burst has a nearly monochromatic spectrum, serving as a distinct characteristic signal that could be detected by radio telescopes. Therefore, we propose that the stimulated decay of collapsing dilute axion stars could be used as a novel standard candle to constrain $H_0$.

What's even more intriguing is that radio signals from the stimulated decay of collapsing axion stars might have already been detected. Axion stars with certain parameters could emit radio bursts that match the characteristics of observed non-repeating FRBs. If some FRBs are indeed generated by collapsing axion stars, this could impose constraints on the axion parameter space, guiding experimental searches for axions. Additionally, collapsing axion stars could provide more accurate measurements of luminosity distance, which would help pinpoint the host galaxies of FRBs. As more FRBs are precisely located, the measurement of the Hubble constant $H_0$ will become more accurate, potentially resolving the Hubble tension. Thus, we propose that FRBs generated by collapsing axion stars could serve as potential standard candles to constrain $H_0$. Axion stars, the ``darkest" matter in the Universe, and FRBs, the ``brightest" radio sources, may together help resolve the Hubble tension.

\section{Acknowledgments}

This work was supported by National Natural Science Foundation of China under Grant No. 11947031 and East China University of Technology Research Foundation for Advanced Talents under Grant No. DHBK2019206.

\end{document}